\documentclass[manuscript, nonacm]{acmart}
\usepackage{graphicx}
\usepackage{float}
\usepackage{booktabs}
\usepackage{multirow}
\usepackage{array}
\usepackage{tikz}
\usetikzlibrary{positioning,arrows}
\usepackage{pgfplots}
\pgfplotsset{compat=1.18}

\title{Beyond Checklist Compliance: A Tale of the Accessibility-Usability Dilemma across Government and Non-Government Web Systems in Bangladesh}

\author{Sanjida Islam Era}
\authornote{Corresponding author.}
\email{era.sanjidaislam@gmail.com}
\affiliation{%
  \department{Department of Computer Science and Engineering}
  \institution{Bangladesh University of Engineering and Technology}
  \city{Dhaka}
  \country{Bangladesh}
}

\author{Tanjeem Azwad Zaman}
\affiliation{%
  \institution{Bangladesh University of Engineering and Technology}
  \department{Department of Computer Science and Engineering}
  \city{Dhaka}
  \country{Bangladesh}
}

\author{Ishika Tarin Ime}
\affiliation{%
  \institution{Bangladesh University of Engineering and Technology}
  \department{Department of Computer Science and Engineering}
  \city{Dhaka}
  \country{Bangladesh}
}

\author{Ashequl Islam Tofa}
\affiliation{%
  \institution{Mawlana Bhashani Science and Technology University}
  \department{Department of Computer Science and Engineering}
  \city{Tangail}
  \country{Bangladesh}
}

\author{Farzin Awsaf Zaman}
\affiliation{%
  \institution{Bangladesh University of Engineering and Technology}
  \department{Department of Computer Science and Engineering}
  \city{Dhaka}
  \country{Bangladesh}
}

\author{A.B.M. Alim al Islam}
\affiliation{%
  \institution{Bangladesh University of Engineering and Technology}
  \department{Department of Computer Science and Engineering}
  \city{Dhaka}
  \country{Bangladesh}
}

\begin{document}

\begin{abstract}
Examining whether technical accessibility measures align with users' usability experiences across web systems in Bangladesh is of utmost significance, given the country's rapidly expanding digital ecosystem. We, therefore, evaluated 207 web systems using WAVE, AChecks, and Squidler and surveyed 103 young adult users. Automated evaluation revealed mixed patterns across government and non-government sectors, with neither consistently performing better across accessibility, functionality, web practices, and performance. In contrast, participants clearly favored non-government platforms for navigation, usability, clarity, and cross-device experience. Qualitative analysis identified three barriers to practical access: system instability disrupting task completion, interaction and navigational friction, and socio-technical barriers involving information access, service continuity, and trust. These findings show that automated accessibility measures capture only part of users' experiences. Accessible web services therefore require attention beyond detectable accessibility problems to reliable task progression, understandable interaction, information access, and continuity across the broader service process.
\end{abstract}


\maketitle

\section{Introduction}

Digital platforms have become essential infrastructure for accessing information and everyday services, making web accessibility critical for equitable participation. This dynamic is particularly significant in Bangladesh, a rapidly digitizing country of over 170 million people where more than half of the population uses the Internet \cite{worldbank2025population,itu2025internet}. Across this expanding digital ecosystem, government services and everyday commercial activities are increasingly supported through online platforms \cite{bangladeshportal2026services,bangladeshbank2025payments}. As reliance on these services grows, however, their expansion raises an important concern: whether the services available online are also accessible and usable enough for people to effectively obtain information and complete the tasks they depend on them for.

Recent evidence suggests that accessibility remains a concern within Bangladesh's growing digital ecosystem. A 2026 UNDP policy brief identified inaccessible online services among the barriers to digital inclusion for persons with disabilities in Bangladesh \cite{undp2026digitalaccessibility}. Recent technical evidence reinforces this concern. A WCAG 2.1 audit of 47 government e-services across 74 websites and 1,534 user-critical pages identified more than 14,000 accessibility violations through automated and expert-led manual evaluation, placing more than half of the evaluated services in the study's lowest accessibility maturity tier \cite{ahsan2025wcag}. A recent national review further highlighted concerns around the accessibility of government digital services, access to assistive technologies, and the implementation of accessibility policies \cite{ssps2025disability}. These concerns are also reflected in users' experiences. A recent study of Bangladeshi hospital websites found difficulties with navigation, system performance, and completing online services \cite{neha2026hospital}. Some participants turned to phone calls, social media, or in-person visits when the websites did not meet their needs \cite{neha2026hospital}. Together, these findings suggest that the growing availability of digital services does not necessarily ensure that people can access and use them effectively.

These challenges extend beyond Bangladesh. A recent large-scale evaluation of 100,000 websites across ten Global South countries documented substantial accessibility barriers and considerable differences across countries and disability-related needs \cite{bhuiyan2025nonwestern}. However, automated evaluation provides only a partial view of accessibility. Existing tools differ in their evaluation capabilities and may consider only a limited set of accessibility criteria \cite{ara2025evaluation,fischer2025coverage}. Moreover, some accessibility requirements cannot be assessed automatically, while user requirements, expert feedback, and semantic aspects may receive insufficient attention in automated evaluation \cite{ara2025framework}. This makes human evaluation important for understanding accessibility beyond what technical measures alone can identify.

Taken together, recent work shows the value of considering both technical evaluation and users' experiences when examining web accessibility. However, these perspectives are often studied separately. Large-scale studies primarily characterize technical accessibility problems across websites \cite{ahsan2025wcag,bhuiyan2025nonwestern}, while recent user-centered work in Bangladesh has examined users' experiences within specific domains such as healthcare \cite{neha2026hospital}. What remains less understood is how measurable website characteristics relate to users' actual experiences across a broader range of government and non-government digital services, and how these experiences relate to information access and institutional trust.

To address this gap, we conducted a mixed-methods study combining automated evaluation of 207 Bangladeshi websites (118 government and 89 non-government) with a survey of 103 young adult users examining a subset of widely used digital services. We further analyzed participants' open-ended responses to understand the barriers they encountered and the improvements they prioritized. Together, these methods allow us to examine technical accessibility at scale alongside users' experiences of interacting with digital services. We address the following research questions:

\vspace{-0.15em}
\begin{itemize}
    \item \textbf{RQ1:} What is the state of accessibility and performance across Bangladesh's government and non-government web platforms, as measured by automated evaluation?
    \item \textbf{RQ2:} What real-world usability and accessibility barriers do young adult users encounter when attempting to complete digital tasks on these platforms?
    \item \textbf{RQ3:} How do these digital experiences relate to users' information access and institutional trust across government and non-government sectors?
\end{itemize}

\vspace{-0.2em}

Our automated evaluation revealed substantial variation across accessibility, web implementation, and functionality measures, with no consistent technical advantage for either sector. Participants reported a clearer preference for non-government platforms, but their accounts also revealed barriers that extended beyond interface usability. System failures, difficult navigation, and interrupted transactions hindered task progression, while incomplete online processes and information gaps sometimes required additional offline steps to obtain services. These experiences show how barriers across digital and offline service processes can limit practical access to information and essential services, while also shaping participants' perceptions of reliability and trust. Building on these findings, this study makes two contributions:
\vspace{-0.49em}
\begin{itemize}
    \item We provide a mixed-methods empirical account of accessibility and usability across Bangladesh's government and non-government web ecosystem, connecting technical evaluation with users' experiences of accessing information and completing digital services.
    \item We identify user-grounded priorities for improving accessibility and service delivery, showing the importance of evaluating complete task and service processes alongside technical accessibility measures.
\end{itemize}
% \vspace{-0.4em}
\section{Related Work}

Digital platforms increasingly shape access to communication, education, banking, commerce, and public services, making accessibility an important concern in everyday digital use. Prior research has examined accessibility across websites, mobile applications, government portals, and educational platforms, using both standards-based assessments and studies of users' experiences~\cite{ahsan2025wcag,pereira2024exploring,huq2026bridging}. We draw on this work to consider how accessibility and usability have been studied across different digital services, where automated evaluations align with or differ from users' experiences, and the broader shift from interface-level accessibility to practical service access.

\subsection{Accessibility Evaluation of Web and Mobile Applications}

Research on web and mobile accessibility continues to identify barriers in navigation, color contrast, alternative text, touch targets, and other interaction requirements~\cite{bhagat2024accessibility,alayed2024saudi,bin2024global}. These problems appear across different application domains. Bhagat et al.~\cite{bhagat2024accessibility} found that assistive applications can introduce difficulties through complex navigation and inconsistent AI responses. Banking applications have similarly shown substantial WCAG compliance problems involving contrast, navigation, and alternative text~\cite{alayed2024saudi}. Across a wider sample, bin Ahsan et al.~\cite{bin2024global} identified recurring accessibility issues in mobile applications from twelve countries.

Recent work has also raised questions about how accessibility is evaluated. Seixas Pereira et al.~\cite{pereira2024exploring} found inconsistencies across assessment methods and limited use of user testing. Accessibility may also depend on the linguistic and cultural setting of an interface. Studies have linked typography, text size, and contrast to accessibility in script-specific contexts~\cite{punsongserm2025enhancing}, while cultural conventions can further shape accessible design~\cite{nourian2025culture}. These considerations are particularly relevant in multilingual contexts, where language presentation, information clarity, and localization may affect users' ability to understand and interact with digital content.

\subsection{Accessibility and Usability of Digital Government and Educational Services}

Government and educational platforms represent particularly important accessibility contexts because they provide access to essential information and services. Previous studies have repeatedly identified accessibility problems within these systems, including missing alternative text, keyboard-navigation barriers, and inadequate semantic structure~\cite{devi2024accessibility,faizin2024indonesia}. Pennathur et al.~\cite{pennathur2024evaluating} evaluated U.S. state public-health websites during the COVID-19 pandemic and found accessibility violations involving non-text content, contrast, and text resizing. Accessibility problems have also been documented across higher-education websites, including in a study of 198 universities in the Gulf region~\cite{fakrudeen2024evaluation}. Accessibility also does not necessarily follow institutional prominence. Abdulloh et al.~\cite{abdulloh2024accessibility} found no clear relationship between university Webometrics rankings and accessibility performance. Rajabli~\cite{rajabliweb} examined accessibility and usability in Massive Open Online Courses (MOOCs), identifying barriers in these online learning environments and highlighting the value of combining multiple forms of accessibility assessment.

Evidence from Bangladesh points to problems that extend beyond individual accessibility violations. Neha et al.~\cite{neha2026hospital} found that hospital websites were affected by slow or non-loading pages, broken links, and difficulties with login and appointment functions. Students using a Bangladeshi public university website likewise reported problems with loading and finding required information~\cite{pranty2025usability}. These findings show how access can break down through problems with the broader service, not only through individual interface elements.

Usability research further suggests that access to digital services depends on how users progress through the interactions required to complete a task. Task-based evaluations of e-government services have therefore examined not only interface characteristics but also how users locate information and move through service processes~\cite{liu2023qingdao}. Research has also connected interaction-related design characteristics with users' perceptions of service quality and their intentions to continue using digital services~\cite{chan2025design}. These studies motivate examining accessibility together with navigation, usability, and task completion rather than treating them as independent characteristics of a service.

\subsection{Automated Accessibility Evaluation and the User Experience}

Automated accessibility tools are widely used because they enable relatively efficient evaluation of large numbers of websites. Tools such as WAVE, AChecks, and Squidler.io can identify violations related to WCAG criteria and reveal recurring technical accessibility problems. However, prior work consistently shows that automated tools differ in what they detect and cannot fully evaluate all accessibility requirements.

Singh et al.~\cite{singh2024accessibility} compared six WCAG evaluation tools across 13 educational websites and found substantial differences in their results, including the detection of alternative-text and text-related violations. Other comparisons have found similar variation in the criteria and accessibility issues covered by different tools~\cite{fischer2025coverage,ara2025evaluation}. Faneer and Elbibas~\cite{faneer2024compliance} combined automated and manual evaluation of Libyan government websites, showing how human assessment can identify issues beyond automated testing.

This gap is also visible when technical evaluation is considered alongside users' experiences. Studies in Argentina and Poland combined accessibility testing with user-centered assessment, capturing barriers that technical measures alone did not fully characterize~\cite{harari2025argentina,zajac2025poland}. Research on Philippine e-government services has similarly emphasized usability as an important component of users' overall experiences with government platforms \cite{philippines2025usability}. The limitations of automated evaluation also affect how accessibility scores should be interpreted. Different tools cover different subsets of accessibility criteria, and some requirements depend on judgments that cannot be reliably automated~\cite{fischer2025coverage,ara2025framework}. Automated results are therefore better understood as evidence of detectable accessibility problems rather than a complete assessment of conformance or accessible use.

In our study, we therefore use multiple automated tools rather than treating the output of a single tool as a comprehensive measure of accessibility. We complement these technical evaluations with users' reported experiences to examine where automated measures align with, and where they differ from, users' experiences of digital services.

\subsection{From Interface Accessibility to Practical Service Access}

Accessibility and usability research increasingly suggests that digital access should be considered at the level of the broader service experience. Drivas and Vraimaki~\cite{drivas2025evaluating}, for example, evaluated 234 museum websites across accessibility, usability, SEO, and performance, finding differences between desktop and mobile experiences. Their work shows the value of considering accessibility alongside other aspects of website quality that shape interaction.

Accessing a service can also require more than interacting with a single webpage. Authentication, information retrieval, form submission, payment, or booking may all be part of the same process, and problems along that process can interrupt access. Buyannemekh et al.~\cite{buyannemekh2026burden}, for example, describe citizens' interactions with government services as processes that may involve websites, telephone communication, and in-person visits, with additional steps imposing further burdens on users. Chan et al.~\cite{chan2021service} similarly describe e-government as a multidimensional service experience shaped by information completeness, accessibility, self-service capability, user support, privacy, and security. These perspectives shift attention from an isolated interface to the broader process users navigate to obtain a service.

Reliability and trust constitute another part of this broader experience. Catalá et al.~\cite{catala2025digitalservices}, studying a digital administrative procedure in Spain, found that reliability and security contributed to whether digital services met users' expectations. Li et al.~\cite{li2023egovernmenttrust} further showed that experiences with government websites can relate to institutional trust through users' perceptions of government integrity, performance, and responsiveness. A meta-analysis of e-government research also highlights the roles of trust, risk, privacy, and security in citizens' use of these services~\cite{gupta2025trust}. In commercial environments, information confidentiality and integrity have similarly been associated with consumers' trusting beliefs toward e-commerce platforms~\cite{singh2024ecommerce}, and broader evidence links online trust with factors such as perceived privacy and service quality~\cite{kim2017trust}.

Taken together, prior work demonstrates substantial progress in identifying technical accessibility barriers, evaluating government and educational websites, and understanding the limitations of automated assessment. However, three gaps remain particularly relevant. First, relatively little work directly examines whether differences observed through multiple automated accessibility measures correspond to differences in users' reported experiences. Second, studies often focus on a specific website, institution, or service category rather than comparing accessibility across a broader government and non-government digital ecosystem. Third, accessibility evaluation frequently centers on interface-level compliance, while users' ability to access a service may also depend on system stability, successful task progression, information availability, service continuity, and trust. We address these gaps by combining automated evaluation of government and non-government websites in Bangladesh with users' quantitative and qualitative accounts of their experiences across these services.

\section{Research Methodology}

We conducted a mixed-methods study combining automated website evaluation with a user survey to examine both technically measurable characteristics of Bangladeshi web platforms and users' experiences of accessing and using digital services~\cite{huq2026bridging,pereira2024exploring}. The research was conducted between 2025 and 2026, encompassing data collection, quantitative analysis of closed-ended survey responses, and qualitative analysis of open-ended responses to identify experienced barriers and user-identified priorities~\cite{nevsky2024lights}. The following sections describe the website evaluation, survey design, and analysis procedures.

\subsection{Website Selection and Categorization}

We purposively selected 207 Bangladeshi websites for automated evaluation, comprising 118 government and 89 non-government platforms. The selection was informed by the Bangladesh National Portal and Bangladeshi web directories covering government and popular websites~\cite{banglasitesGovernment,banglasitesPopular,bangladeshNationalPortal}. We prioritized websites providing essential information, common online activities, and service-related tasks, aiming to capture variation in website purpose and interaction requirements rather than a statistically representative sample~\cite{Hambley2025Population}.

The government group included national portals, ministries, agencies, and public service platforms, while the non-government group covered news and media, e-commerce, financial services, telecommunications, travel, education, and employment.

\subsection{Automated Website Evaluation}

We evaluated the 207 selected websites using three automated tools: Web Accessibility Evaluation Tool (WAVE), AChecks, and Squidler. Together, the tools captured different aspects of accessibility, web implementation practices, functionality, and performance. Websites that blocked automated evaluation, failed to load, or were otherwise unavailable were recorded as missing and excluded from the corresponding tool-specific analysis. The resulting coverage was 179 websites for WAVE (113 government, 66 non-government), 190 for AChecks (111 government, 79 non-government), and 156 for Squidler (75 government, 81 non-government).

\begin{table*}[h!]
\centering
\caption{Automated evaluation tools, extracted metrics, and primary evaluation focus.}
\label{tab:automated_tools}
\begin{tabular}{p{1.7cm} p{4.4cm} p{7.3cm}}
    \toprule
    \textbf{Tool} & \textbf{Metrics Analyzed} & \textbf{Evaluation Focus} \\
    \midrule

    \textbf{WAVE} &
    Errors, Contrast Errors, Alerts, Features, Structural Elements, ARIA &
    Accessibility barriers and accessibility-supporting page structures related to WCAG 2.2 criteria. \\

    \addlinespace

    \textbf{AChecks} &
    Accessibility, Best Practices, SEO, Speed &
    Accessibility, general web implementation practices, and performance. \\

    \addlinespace

    \textbf{Squidler} &
    Pages, Actions, Functionality, Accessibility, Language &
    Website-level accessibility, functionality, interaction, and language-related issues. \\

    \bottomrule
\end{tabular}
\end{table*}

% \vspace{-0.5em}
\subsubsection{WAVE}

WAVE was applied to each website's homepage to identify accessibility-related issues using checks associated with WCAG 2.2. We extracted the six categories summarized in Table~\ref{tab:automated_tools}. Errors and Contrast Errors identify detected accessibility barriers, while Alerts flag potential issues requiring human review. Features identify accessibility-supporting elements, Structural Elements describe the organization of page content, and Accessible Rich Internet Applications (ARIA) identifies attributes supporting assistive technologies~\cite{WebAIMWAVEDocumentation}.

% \vspace{-0.5em}
\subsubsection{AChecks}

AChecks was used to evaluate accessibility, web implementation practices, and performance through its Lighthouse-based assessment, which checks rendered webpages against WCAG 2.1 AA~\cite{AChecksDocumentation}. We analyzed the four scores summarized in Table~\ref{tab:automated_tools}, each reported on a 0--100 scale, with higher values indicating better results. Accessibility assesses automated accessibility checks, Best Practices evaluates implementation quality and security-related practices, Search Engine Optimization (SEO) assesses search-engine discoverability, and Speed measures loading and interaction performance.

% \vspace{-0.5em}
\subsubsection{Squidler}

Squidler extended the evaluation through automated browser-based navigation and interaction across website pages. During our data-collection period, the platform reported five metrics: Pages, Actions, Functionality, Accessibility, and Language. Pages and Actions described the scope of pages and interactions explored during the scan, while Functionality, Accessibility, and Language represented the corresponding issue counts reported by the platform. Lower issue counts indicated fewer detected problems within the scope of the automated evaluation. We successfully evaluated 156 websites (75 government and 81 non-government) using Squidler. Further details on the evaluation process and metric definitions used during the study are provided in Appendix~\ref{sec:appendix_squidler}.

\subsection{User Survey}

To complement the automated evaluation, we conducted a structured user survey to capture users' experiences with the accessibility, usability, performance, and perceived security of Bangladeshi web platforms. The questionnaire included both closed-ended measures and open-ended questions, allowing participants to report barriers and experiences that may not be captured through automated evaluation.

\subsubsection{Platform Selection for Survey}

For the closed-ended comparative evaluations, we selected 20 commonly used platforms (10 government and 10 non-government) from the broader corpus (Table~\ref{tab:survey_websites}). The selection aimed to represent different digital service categories, including civil registration, public services, transportation, e-commerce, employment, travel, and education, with several selected commercial platforms also documented in prior research and national digital commerce reports~\cite{Faria2024Ecommerce,USITABangladeshEcommerce}. The open-ended questions were not restricted to these platforms, allowing participants to report barriers encountered across other Bangladeshi websites.

\begin{table*}[h!]
\centering
\small
\caption{Government and non-government platforms included in the closed-ended survey evaluation.}
\label{tab:survey_websites}
\begin{tabular}{c l c l}
\toprule
\textbf{\#} & \textbf{Government Platform} &
\textbf{\#} & \textbf{Non-Government Platform} \\
\midrule
1  & E-Passport Portal            & 11 & Daraz (E-commerce) \\
2  & E-Visa Portal                & 12 & Sheba (Service Marketplace) \\
3  & E-TIN Registration           & 13 & GoZayaan (Travel Booking) \\
4  & Bangladesh Railway Portal    & 14 & Rokomari (Bookstore) \\
5  & Election Commission (NIDW)   & 15 & Chaldal (Grocery) \\
6  & Digital Land Management      & 16 & Aarong (Retail/Lifestyle) \\
7  & Birth Registration (BDRIS)   & 17 & Bdjobs (Employment) \\
8  & National Web Portal          & 18 & Bikroy (Classifieds) \\
9  & Ministry of Education        & 19 & Shajgoj (Beauty/Wellness) \\
10 & Ministry of Foreign Affairs  & 20 & Shikho (E-Learning) \\
\bottomrule
\end{tabular}
\end{table*}

\subsubsection{Recruitment and Participants}
Participants were recruited through social-media platforms using convenience sampling, and the survey was administered online through Google Forms. The survey included $N=103$ participants, predominantly young adult internet users. Most were aged 20--35 ($n=98$, 95.1\%); 77 (74.8\%) identified as male and 26 (25.2\%) as female. Students formed the largest occupational group ($n=83$, 80.6\%), and 59 participants reported a Computer Science and Engineering background. Most held or were pursuing a university-level degree ($n=74$, 71.8\%).

Participants generally reported substantial experience using digital technologies: 53 (51.5\%) identified as advanced internet users and 43 (41.7\%) as intermediate users. Most accessed websites through desktop or laptop computers ($n=98$, 95.1\%) and smartphones ($n=93$, 90.3\%); participants could report more than one device. Participants were assigned anonymized identifiers (P001--P103) for attributing open-ended responses.

\subsubsection{Survey Instrument}

The questionnaire was designed to complement the automated evaluation by capturing participants' reported experiences of accessing and using Bangladeshi web platforms. Following the distinction between technical accessibility evaluation and user-experienced barriers emphasized in prior HCI research~\cite{pereira2024exploring,huq2026bridging}, we organized the closed-ended questions into four areas. These questions referred to the 20 selected platforms described in Table~\ref{tab:survey_websites}.

The usability and task completion area asked participants to rate navigation ease and report whether they could complete intended activities, such as booking tickets, submitting forms, or making purchases. User experience questions examined how easily participants could find information or services, understand instructions and labels, and use websites across different devices, with parallel ratings for government and non-government platforms. The accessibility area asked whether participants noticed features such as larger text, contrasting colors, or simplified navigation, and which features they found helpful. Performance and security questions captured perceived loading speed, experiences of crashes or errors, confidence in providing personal information, and encounters with suspicious or security-related activity. The questionnaire used five-point rating scales, categorical responses, and platform-selection questions, as appropriate to each item.

Four open-ended questions separately invited participants to describe website-specific usability problems, crashes or errors, security-related issues, and suggestions for improving accessibility and usability. These responses formed the basis of the qualitative analyses described in Sections~\ref{sec:thematic_analysis} and~\ref{sec:user_priorities}.

\subsection{Quantitative Data Analysis}

We analyzed two sources of quantitative data: automated website measurements and closed-ended survey responses. The analysis described accessibility, performance, and user experience patterns and compared these patterns across government and non-government platforms.

\vspace{-0.8em}
\subsubsection{Automated Evaluation Data}
Measurements from WAVE, AChecks, and Squidler were analyzed separately for government and non-government websites, using only websites successfully evaluated by each tool. For each metric, we summarized sector-level distributions using means, medians, and standard deviations. We compared the resulting distributions to examine patterns in detected accessibility issues, web implementation practices, performance, and functionality~\cite{Fernandes2023Accessibility,Frandell2026Breaking}. The comparisons addressed RQ1 without combining metrics from different tools into a single accessibility score.

\vspace{-0.8em}
\subsubsection{Closed-Ended Survey Data}
Closed-ended survey responses were summarized using counts, percentages, and five-point rating distributions, preserving the response patterns of individual survey items~\cite{Sullivan2013Likert,Boone2012Likert}. For the 20 selected platforms, we examined platform-level responses to characterize participants' reported experiences with individual services. We also compared parallel ratings and other sector-level responses to examine patterns across government and non-government platforms.

\subsection{Thematic Analysis of Experienced Digital Barriers}
\label{sec:thematic_analysis}

To investigate the accessibility and usability barriers reported by participants (RQ2), we conducted an inductive thematic analysis following Braun and Clarke's approach~\cite{BraunClarke2006,Terry2017Thematic}. We used open coding and iterative category development to organize reported issues into higher-level patterns before synthesizing the final themes.

The first author conducted the initial coding and developed the coding structure. A coauthor with prior experience in thematic analysis reviewed the evolving codes, categories, and themes against the original responses. The authors discussed ambiguous or overlapping interpretations and refined the structure collaboratively~\cite{huq2026bridging,Quaium2023Hajj}. We coded the problems participants reported without assuming technical causes that could not be verified from their responses.

\vspace{-0.8em}
\paragraph{Open Coding and Refinement.}
We repeatedly reviewed the responses and generated initial codes describing specific reported barriers. Related or overlapping codes were consolidated, and their definitions were refined through comparison with the original responses. Established HCI concepts informed the interpretation where relevant, while the coding remained grounded in participants' accounts. This process yielded 54 refined open codes.

\vspace{-0.8em}
\paragraph{Category Development.}
We grouped related open codes into broader categories, examining similarities and distinctions across reported experiences. The categories were iteratively reviewed and refined, resulting in 10 axial categories.

\vspace{-0.8em}
\paragraph{Thematic Synthesis.}
We examined relationships among the categories to develop overarching patterns of meaning. Candidate themes were reviewed against the underlying codes and original responses, refined to distinguish their scope, and assigned descriptive names. This process resulted in three themes, presented with their supporting categories and participant accounts in Section~\ref{sec:qualitative_findings}.

\subsection{Qualitative Content Analysis of User-Identified Priorities}
\label{sec:user_priorities}

We conducted a separate inductive qualitative content analysis of the open-ended recommendation responses to identify participants' suggested improvements. The analysis focused on proposed interventions rather than the experienced barriers examined in RQ2. We developed categories from the responses rather than applying a predefined coding framework~\cite{HsiehShannon2005,EloKyngas2008}.

\vspace{-0.5em}
\paragraph{Recommendation Consolidation.}
We extracted 27 raw suggestions and merged those expressing the same underlying recommendation, such as ``improve loading speed'' and ``websites should be faster.'' This resulted in 23 distinct recommendation codes.

\vspace{-0.5em}
\paragraph{Priority Categorization.}
We grouped the 23 codes into five higher-order categories based on their target intervention areas. The resulting priorities and supporting recommendations are presented in Section~\ref{sec:user_identified_priorities}.
\vspace{-0.5em}

\subsection{Cross-Method Synthesis}
\label{sec:cross_method}

To address RQ3, we integrated the automated evaluation, quantitative survey results, and qualitative findings to examine how digital experiences relate to information access and institutional trust. We compared findings across methods to identify areas of convergence, divergence, and complementarity, following established mixed-methods integration principles~\cite{Fetters2013Integration}.

We considered survey measures of task completion, navigation, and performance alongside the technical evaluation and Themes 1 and 2 to examine how accessibility and usability barriers related to access to information and digital services. Measures of confidence in providing personal information and sector-specific security preferences were examined alongside Theme 3, which captured experiences concerning service fulfillment, information clarity, transparency, and trust. These comparisons informed our interpretation of how technical conditions and reported experiences jointly shaped access and trust, without treating automated scores as direct measures of usability or inferring causal relationships from the survey. As automated tools and survey captured different levels of the digital experience, we used this synthesis to examine broader patterns of convergence and divergence instead of assuming direct correspondence between them.

\section{Results}

We present the findings in four parts. We first characterize the technical accessibility, web practices, functionality, and performance of the evaluated websites, then examine users' reported experiences across government and non-government platforms. We subsequently use participants' open-ended responses to explain the barriers underlying these experiences and conclude with the improvements participants prioritized.

\subsection{Automated Evaluation of the Web Ecosystem}
\label{sec:automated_results}

The automated evaluation addressed RQ1 by characterizing technical accessibility, web practices, functionality, and performance across the Bangladeshi web ecosystem. Table~\ref{tab:automated_consolidated} summarizes the results by tool and sector. Across the evaluated websites, the direction of government--non-government differences varied across tools and metrics, with no sector showing a consistent technical advantage.

\begin{table*}[h!]
\centering
\caption{Consolidated automated evaluation metrics for government, non-government, and overall website samples.}
\label{tab:automated_consolidated}
\resizebox{\textwidth}{!}{
\begin{tabular}{llrrrrrrrrr}
\toprule
\multirow{2}{*}{Tool} & \multirow{2}{*}{Metric}
& \multicolumn{3}{c}{Government}
& \multicolumn{3}{c}{Non-Government}
& \multicolumn{3}{c}{Overall} \\
\cmidrule(lr){3-5}\cmidrule(lr){6-8}\cmidrule(lr){9-11}
& & Mean & Med & SD & Mean & Med & SD & Mean & Med & SD \\
\midrule

\multirow{6}{*}{\textbf{WAVE}}
& Errors              & 25.56 & 26.0 & 7.89  & 46.67 & 21.0 & 72.37  & 33.34 & 26.0 & 45.34 \\
& Contrast Issues     & 22.10 & 22.0 & 13.32 & 43.38 & 19.5 & 80.17  & 29.94 & 22.0 & 50.64 \\
& Alerts              & 134.47 & 146.0 & 53.62 & 99.91 & 61.0 & 108.88 & 121.73 & 134.0 & 80.11 \\
& Features            & 29.01 & 28.0 & 11.25 & 59.30 & 31.0 & 73.15  & 40.18 & 28.0 & 47.42 \\
& Structural Elements & 83.16 & 86.0 & 28.35 & 107.71 & 67.5 & 105.50 & 92.21 & 84.0 & 68.64 \\
& ARIA Usage          & 73.69 & 76.0 & 52.74 & 98.35 & 37.0 & 165.16 & 82.78 & 67.0 & 108.87 \\

\midrule

\multirow{4}{*}{\textbf{AChecks}}
& Accessibility       & 74.48 & 75.0 & 5.83  & 79.53 & 80.0 & 11.63 & 76.58 & 76.0 & 9.05 \\
& Best Practice       & 72.16 & 70.0 & 7.40  & 88.46 & 93.0 & 11.42 & 78.94 & 70.0 & 12.27 \\
& SEO                 & 76.12 & 75.0 & 5.71  & 83.18 & 85.0 & 16.94 & 79.05 & 75.0 & 12.23 \\
& Speed               & 44.13 & 44.0 & 14.21 & 42.57 & 40.0 & 22.60 & 43.48 & 42.0 & 18.14 \\

\midrule

\multirow{5}{*}{\textbf{Squidler}}
& Accessibility       & 10.63 & 4.0  & 13.55 & 18.38 & 14.0 & 17.15 & 14.65 & 9.5  & 15.96 \\
& Functionality       & 16.12 & 12.0 & 16.07 & 3.52  & 2.0  & 7.62  & 9.58  & 4.5  & 13.90 \\
& Language            & 1.28  & 0.0  & 2.54  & 2.99  & 0.0  & 5.60  & 2.17  & 0.0  & 4.47 \\
& Pages               & 4.39  & 2.0  & 6.46  & 9.62  & 9.0  & 6.47  & 7.10  & 5.0  & 6.95 \\
& Actions             & 43.76 & 46.0 & 25.93 & 42.95 & 41.0 & 24.43 & 43.34 & 44.0 & 25.08 \\

\bottomrule
\end{tabular}}
\end{table*}

% \vspace{-0.6em}
\paragraph{Accessibility Measures Varied across Tools.}
Government websites recorded fewer WAVE Errors (25.56 vs.\ 46.67) and Contrast Issues (22.10 vs.\ 43.38), as well as fewer Squidler Accessibility issues (10.63 vs.\ 18.38). In contrast, non-government websites achieved a higher mean AChecks Accessibility score (79.53 vs.\ 74.48). The sector comparison therefore changed direction across tools, reflecting differences in what each tool measures and in the subsets of websites successfully evaluated. The WAVE results also showed substantial variation among non-government websites: Errors had a mean of 46.67 but a median of 21.0 ($SD=72.37$), indicating that a smaller set of websites contributed disproportionately to the sector mean.

% \vspace{-0.6em}
\paragraph{Web Practices, Functionality, and Performance Showed Different Sector Patterns.}
Non-government websites achieved higher AChecks Best Practice (88.46 vs.\ 72.16) and SEO scores (83.18 vs.\ 76.12). They also had higher mean counts of WAVE Features, Structural Elements, and ARIA Usage, although these measures describe detected webpage characteristics rather than directly indicating better accessibility. Squidler identified substantially more Functionality issues on government websites (16.12 vs.\ 3.52), while fewer Pages were reached during evaluation (4.39 vs.\ 9.62). Actions were similar across sectors (43.76 vs.\ 42.95). AChecks Speed was slightly higher for government websites (44.13 vs.\ 42.57), although both sectors had comparatively low mean scores.

\vspace{0.5em}
Taken together, the automated evaluation revealed no uniform technical advantage for either sector. Government websites showed fewer detected issues on selected WAVE and Squidler accessibility measures, whereas non-government websites performed more favorably on AChecks Accessibility, Best Practice, and SEO and showed fewer Squidler Functionality issues. These cross-tool differences reinforce the need to interpret automated accessibility measures separately rather than as interchangeable indicators of overall accessibility~\cite{fischer2025coverage}. The following section examines whether these mixed technical patterns are reflected in users' reported experiences.

\subsection{User-Reported Experiences Across the Web Ecosystem}
\label{sec:survey_results}
The closed-ended survey responses revealed a clearer sector-level pattern than the automated evaluation, with participants generally reporting more favorable experiences on non-government platforms.

\subsubsection{Navigation and Task Completion}
Participants strongly favored non-government websites for navigation, with 82.5\% (85/103) identifying them as easier to navigate, compared with 2.9\% (3/103) selecting government websites. Task-completion experiences were also more favorable on frequently used non-government platforms: 49 participants completed their intended tasks on Daraz without issues and 22 with only minor difficulty, while the corresponding counts for Rokomari were 45 and 15. Government services showed more mixed experiences; for Bangladesh Railway, only 26 participants reported completing their tasks without issues, while 37 experienced minor or moderate difficulty.

\subsubsection{Ease of Use, Clarity, and Cross-Device Experience}

Participants rated non-government websites more favorably across all three experience measures (Figure~\ref{fig:likert_distribution}). Ratings of 4--5 were substantially more common for non-government than government websites for ease of use (81.5\% vs.\ 22.3\%), clarity of instructions (83.5\% vs.\ 21.4\%), and cross-device experience (79.6\% vs.\ 14.6\%). The largest contrast appeared in cross-device experience, where 46.6\% of government ratings were 1--2, compared with only 2.0\% for non-government websites. When asked which platform provided the best overall experience within each sector, Bangladesh Railway (30.1\%) and E-Passport (27.2\%) led among government websites, while Daraz (30.1\%) and Rokomari (20.4\%) led among non-government websites.

\begin{figure*}[h!]
\centering
\begin{tikzpicture}

% Diverging Likert palette
\definecolor{likert1}{HTML}{D7191C}
\definecolor{likert2}{HTML}{FDAE61}
\definecolor{likert3}{HTML}{E0E0E0}
\definecolor{likert4}{HTML}{ABD9E9}
\definecolor{likert5}{HTML}{2C7BB6}

\begin{axis}[
    xbar stacked,
    width=0.84\textwidth,
    height=6.6cm,
    bar width=10pt,
    xmin=0,
    xmax=100,
    ymin=0.4,
    ymax=6.8,
    clip=false,
    xlabel={\textbf{Percentage of Responses (\%)}},
    xlabel style={font=\small},
    xmajorgrids=true,
    grid style={line width=0.5pt, draw=gray!20},
    axis x line*=bottom,
    axis y line=none,
    tick label style={font=\footnotesize},
    ytick=\empty,
    legend style={
        at={(0.5,1.12)},
        anchor=south,
        legend columns=5,
        font=\footnotesize,
        draw=none,
        /tikz/every even column/.append style={column sep=0.35cm}
    }
]

% Rating 1
\addplot[
    fill=likert1,
    draw=black!70,
    line width=0.4pt
] coordinates {
    (1.0,1.0)   % Cross-device: Non-government
    (16.5,1.7)  % Cross-device: Government
    (1.0,3.1)   % Clarity: Non-government
    (9.7,3.8)   % Clarity: Government
    (1.0,5.2)   % Ease: Non-government
    (4.9,5.9)   % Ease: Government
};
\addlegendentry{1}

% Rating 2
\addplot[
    fill=likert2,
    draw=black!70,
    line width=0.4pt
] coordinates {
    (1.0,1.0)
    (30.1,1.7)
    (0.0,3.1)
    (26.2,3.8)
    (1.0,5.2)
    (34.0,5.9)
};
\addlegendentry{2}

% Rating 3
\addplot[
    fill=likert3,
    draw=black!70,
    line width=0.4pt
] coordinates {
    (18.4,1.0)
    (38.8,1.7)
    (15.5,3.1)
    (42.7,3.8)
    (16.5,5.2)
    (38.8,5.9)
};
\addlegendentry{3}

% Rating 4
\addplot[
    fill=likert4,
    draw=black!70,
    line width=0.4pt
] coordinates {
    (58.3,1.0)
    (11.7,1.7)
    (64.1,3.1)
    (16.5,3.8)
    (62.1,5.2)
    (19.4,5.9)
};
\addlegendentry{4}

% Rating 5
\addplot[
    fill=likert5,
    draw=black!70,
    line width=0.4pt
] coordinates {
    (21.3,1.0)
    (2.9,1.7)
    (19.4,3.1)
    (4.9,3.8)
    (19.4,5.2)
    (2.9,5.9)
};
\addlegendentry{5}

% Ease of use
\node[
    anchor=east,
    font=\small\bfseries
] at (axis cs:-0.25,6.35) {Ease of use};

\node[
    anchor=east,
    font=\footnotesize
] at (axis cs:-0.25,5.9) {Government};

\node[
    anchor=east,
    font=\footnotesize
] at (axis cs:-0.25,5.2) {Non-government};

% Clarity of instructions
\node[
    anchor=east,
    font=\small\bfseries
] at (axis cs:-0.25,4.25) {Clarity of instructions};

\node[
    anchor=east,
    font=\footnotesize
] at (axis cs:-0.25,3.8) {Government};

\node[
    anchor=east,
    font=\footnotesize
] at (axis cs:-0.25,3.1) {Non-government};

% Cross-device experience
\node[
    anchor=east,
    font=\small\bfseries
] at (axis cs:-0.25,2.15) {Cross-device experience};

\node[
    anchor=east,
    font=\footnotesize
] at (axis cs:-0.25,1.7) {Government};

\node[
    anchor=east,
    font=\footnotesize
] at (axis cs:-0.25,1.0) {Non-government};

\end{axis}
\end{tikzpicture}

\caption{Ratings of ease of use, clarity of instructions, and cross-device experience for government and non-government websites ($N=103$). Higher ratings indicate more favorable experiences.}
\label{fig:likert_distribution}
\Description{Stacked bar chart comparing ratings of government and
    non-government websites. The horizontal axis shows the percentage of
    responses from 0 to 100, and the bars show ratings from 1 to 5 for ease
    of use, clarity of instructions, and cross-device experience.
    Non-government websites receive substantially more ratings of 4 and 5
    across all three measures, while government websites receive more ratings
    of 1, 2, and 3.}
\end{figure*}

\subsubsection{Perceived Accessibility}
Participants' overall judgments of accessibility and usability strongly favored non-government platforms, selected by 79.6\% (82/103) compared with 1.9\% (2/103) for government platforms. Accessibility-supporting features were moderately noticeable, with 3/5 being the most common visibility rating. Participants most frequently identified simple and clear navigation, search functionality, mobile responsiveness, and quick loading times as features that made websites easier to use (Figure~\ref{fig:survey_usability_features}), while text resizing, high-contrast mode, and keyboard navigation were selected less frequently.

\begin{figure*}[h!]
\centering
\begin{tikzpicture}
\begin{axis}[
    ybar,
    bar width=13pt,
    width=0.98\textwidth,
    height=7.2cm,
    ymin=0, ymax=80,
    ylabel={\textbf{Number of Responses ($N = 103$)}},
    ylabel style={font=\small},
    % Remove top and right box outlines/ticks
    axis x line*=bottom,
    axis y line*=left,
    axis line style={draw=black!70},
    % Soft horizontal gridlines behind bars
    ymajorgrids=true,
    grid style={line width=0.5pt, draw=gray!20},
    symbolic x coords={
        {Text resizing options},
        {High contrast mode},
        {Keyboard navigation},
        {Simple and clear navigation menu},
        {Search functionality},
        {Mobile responsiveness},
        {Readable fonts and appropriate font sizes},
        {Clear labels or tooltips},
        {Quick loading times},
        {None of the above}
    },
    xtick=data,
    x tick label style={rotate=40, anchor=east, font=\footnotesize},
    y tick label style={font=\small},
    % Value badges above each bar
    nodes near coords,
    every node near coord/.append style={
        font=\scriptsize\bfseries,
        fill=white,
        draw=black!60,
        rounded corners=2pt,
        inner sep=1.8pt,
        yshift=2pt
    },
    enlarge x limits=0.06
]
\addplot[
    fill={rgb,255:red,90;green,180;blue,172}, % Clean Teal matching sample
    draw=black!85,
    line width=0.7pt
] coordinates {
    ({Text resizing options}, 19)
    ({High contrast mode}, 20)
    ({Keyboard navigation}, 12)
    ({Simple and clear navigation menu}, 70)
    ({Search functionality}, 66)
    ({Mobile responsiveness}, 52)
    ({Readable fonts and appropriate font sizes}, 39)
    ({Clear labels or tooltips}, 35)
    ({Quick loading times}, 46)
    ({None of the above}, 6)
};
\end{axis}
\end{tikzpicture}
\caption{Features identified by respondents as contributing most to website usability ($N = 103$), maintaining the original questionnaire sequence.}
\label{fig:survey_usability_features}
 \Description{Bar chart showing website features identified by respondents
    as contributing to usability. The vertical axis shows the number of
    responses from 0 to 80. Simple and clear navigation received 70 responses,
    search functionality 66, mobile responsiveness 52, quick loading times 46,
    readable fonts and appropriate font sizes 39, clear labels or tooltips 35,
    high contrast mode 20, text resizing options 19, keyboard navigation 12,
    and none of the above 6.}
\end{figure*}

% \vspace{-2em}
\subsubsection{Performance and Perceived Security}

Performance concerns were common across the ecosystem: loading speed was most frequently rated as average, while 47.6\% (49/103) reported experiencing website crashes or errors, with a further 27.2\% (28/103) indicating that they may have experienced such problems. Confidence in providing personal information was concentrated in the middle and lower ratings. Perceived security favored non-government websites, selected by 37.9\% (39/103), compared with 19.4\% (20/103) for government websites, while 26.2\% (27/103) selected neither.

Overall, participants reported more favorable experiences with non-government platforms across several measures; the following qualitative findings examine the barriers underlying these experiences.

% --------------------------------new thematic analysis-------------------------
\subsection{Experienced Accessibility and Usability Barriers}
\label{sec:qualitative_findings}

The open-ended responses provided a more detailed account of the barriers participants encountered while using digital services. As described in Section~\ref{sec:thematic_analysis}, 54 refined open codes were organized into 10 axial categories and synthesized into three overarching themes (Table~\ref{tab:qualitative_themes}). The complete codebook is provided in Appendix~\ref{app:qualitative_codebook}.

\begin{table*}[h!]
\centering
\caption{Summary of themes identified through thematic analysis, with descriptions, axial categories, and illustrative participant responses.}
\label{tab:qualitative_themes}
\small
\begin{tabular}{
    p{0.16\textwidth}
    p{0.17\textwidth}
    p{0.20\textwidth}
    p{0.31\textwidth}
}
\toprule
\textbf{Theme} & \textbf{Description} & \textbf{Axial Category} & \textbf{Illustrative Responses} \\
\midrule

% ==================== THEME 1 ====================
\multirow{10}{0.16\textwidth}{\textbf{System Instability and Barriers to Digital Task Completion}} &
\multirow{10}{0.17\textwidth}{Breakdowns that interrupt users' progress through digital tasks or prevent successful completion.} &
\textbf{System Performance, Stability \& Availability} &
``too slow and most of the time doesn't respond'' (P047) \\
\addlinespace
\cline{3-4}
\addlinespace
& & \textbf{Authentication \& Access Barriers} &
``verification code'' arrived late (P066) \\
\addlinespace
\cline{3-4}
\addlinespace
& & \textbf{Task \& Transaction Workflow Failures} &
``without completing full process it deducted [money] from account'' (P028); ``Payment gateway did not load multiple tries'' (P045) \\
\addlinespace
\cline{3-4}
\addlinespace
& & \textbf{System State \& Error Handling Deficits} &
``going back requires login from beginning'' (P030); ``booked seats are shown as available'' (P024) \\
\midrule

% ==================== THEME 2 ====================
\multirow{9}{0.16\textwidth}{\textbf{Interaction Cost and Navigational Friction}} &
\multirow{9}{0.17\textwidth}{Friction that makes interfaces harder to understand, navigate, or use consistently across platforms and devices.} &
\textbf{Interface Design \& Interaction Friction} &
``Flashy, Unorganized, Cluttered UI'' (P082) \\
\addlinespace
\cline{3-4}
\addlinespace
& & \textbf{Navigational \& Wayfinding Friction} &
``info scattered around'' (P037); ``search algorithm is not up to the mark'' (P038) \\
\addlinespace
\cline{3-4}
\addlinespace
& & \textbf{Cross-Platform \& Device Inconsistency} &
``tickets did not load in the website, but noticed the abundance of tickets in mobile app'' (P004) \\
\midrule

% ==================== THEME 3 ====================
\multirow{9}{0.16\textwidth}{\textbf{Socio-Technical Barriers to Information Access and Service Trust}} &
\multirow{9}{0.17\textwidth}{Barriers that limit whether users can obtain and depend on the information and services they need.} &
\textbf{Service Fulfillment \& Digital Discontinuity} &
``corrections cannot be done online'' and required an in-person visit (P023) \\
\addlinespace
\cline{3-4}
\addlinespace
& & \textbf{Information Clarity \& Visibility Deficits} &
``unclear instructions'' (P014); order tracking did not work correctly (P099) \\
\addlinespace
\cline{3-4}
\addlinespace
& & \textbf{Systemic Trust \& Transparency Deficits} &
``chance of being scammed'' (P025); participant-reported concerns about additional payment for service completion (P023) \\
\bottomrule
\end{tabular}
\end{table*}

% \vspace{-0.5em}
\subsubsection{System Instability and Barriers to Digital Task Completion}
\label{sec:theme1}

System instability became consequential when it prevented participants from reliably progressing through digital tasks. Some participants encountered problems before a task could meaningfully begin. P001 reported that the E-TIN website ``did not load properly for a long time,'' while P038 experienced a Bangladesh Railway website crash ``during sign up.'' Access could remain difficult even after a service loaded: P018 encountered an ``OTP issue'' on the Railway portal, while other accounts described delayed verification codes. In a more persistent case, P030 described a Birth Registration account that had been disqualified and remained unresolved for four years. These experiences show how system availability and access mechanisms could obstruct users before they reached or fully entered the service they intended to use.

More consequential breakdowns emerged within multi-step tasks, particularly transactions. P021 described the E-Passport Portal as having an ``extremely faulty online transaction system,'' while P028 reported that a Railway transaction deducted money ``without completing [the] full process,'' although the amount was later refunded. P025 became ``stuck at booking a seat'' after a transaction failed and subsequently found the seat marked as booked, requiring them to wait before attempting another purchase. P045 similarly reported multiple attempts in which the Railway payment gateway did not load. In these cases, an initial transaction failure could therefore extend beyond a single unsuccessful action, leaving users unable to continue or uncertain about the state of an ongoing task.

Problems with preserving progress and recovering from failures appeared elsewhere as well. P030 reported that navigating back on the Railway portal required logging in ``from beginning,'' effectively losing progress, while P098 could not submit an E-TIN registration application and P103 encountered a ``runtime exception'' during registration. These experiences show how failures could leave users with lost progress or unresolved tasks rather than a straightforward opportunity to retry.

Across these accounts, the barrier was not any single crash, delay, or failed interaction, but the inability to rely on a digital process to progress from access to successful completion.

% =========================================================
% THEME 2
% =========================================================

% \vspace{-0.5em}
\subsubsection{Interaction Cost and Navigational Friction}
\label{sec:theme2}

Even when a service remained available, participants described interface problems that made interaction unnecessarily difficult. P074 found Chaldal ``overwhelming'' and its location selection confusing, while P039 reported that selecting a product on Aarong repeatedly opened a new tab. P045 similarly described the E-TIN and Election Commission interfaces as unintuitive, with unclear navigation on the latter. Other participants pointed to clutter and competing content; P069 reported that advertisements loaded before the intended content. These experiences show how interface design could increase interaction effort without necessarily making the underlying service unavailable.

Finding the intended information or action created further friction. P025 described difficulty navigating numerous Daraz listings for the same product across different sellers and prices, complicating product selection. P038 criticized the organization of content on the Ministry of Education website alongside ineffective search on Rokomari. These accounts show how poor information organization and discoverability could require users to spend additional effort determining where relevant content or actions were located.

The experience could also depend on the platform or device through which a service was accessed. P072 reported that books could not be downloaded from the Ministry of Education website on Android, while P004 found Railway tickets unavailable on the website despite their availability in the mobile application. Functionality available through one access point therefore did not always translate into equivalent access through another.

Across these accounts, interaction barriers arose not simply from whether functionality existed, but from the effort required to find, understand, and use it consistently. This also helps contextualize participants' preference for clear navigation, effective search, and mobile responsiveness in the closed-ended survey.

% =========================================================
% THEME 3
% =========================================================
% \vspace{-0.5em}
\subsubsection{Socio-Technical Barriers to Information Access and Service Trust}
\label{sec:theme3}

Accessing a digital platform did not always mean that participants could obtain the underlying service digitally. P023 described a certificate-correction process that ultimately required visiting their school or college in person, with corrections involving different education boards requiring separate applications. The same participant described being unable to find an expected Election Commission update online and instead having to check offline. These experiences show how a digital entry point could still lead to fragmented administrative processes or offline dependencies before the intended service was fulfilled.

Other participants encountered gaps in the information needed to understand or follow a service. P020 described instructions as ``mostly unclear,'' while P099 reported that an expected delivery time was not displayed and that order tracking did not work correctly. P072 similarly reported being unable to find intended content on the Ministry of Education website. Such gaps could leave participants without the information needed to understand what to do next or determine the status of an ongoing service.

Some accounts also raised concerns about whether platforms could be relied upon. P023 described additional unofficial costs associated with completing a public administrative service, while P025 described Daraz as having “less reliability” and a “chance of being scammed.” These concerns align with the broader survey findings on perceived security, suggesting that dependable content and services also shaped users' trust in digital platforms.

Across these accounts, practical digital access depended not only on reaching and using an interface, but also on obtaining the information and services users needed through processes they could depend on.

% \vspace{-0.5em}
\subsection{User-Identified Improvement Priorities}
\label{sec:user_identified_priorities}

Participants were also asked to suggest ways to improve website usability and accessibility. As described in Section~\ref{sec:user_priorities}, 27 raw suggestions were consolidated into 23 distinct recommendation codes and organized into five higher-order improvement priorities. Table~\ref{tab:improvement_priorities} presents the complete mapping.

\begin{table*}[h!]
\centering
\caption{User-identified improvement priorities and their 23 recommendation codes.}
\label{tab:improvement_priorities}
\small
\begin{tabular}{p{0.30\textwidth} p{0.64\textwidth}}
\toprule
\textbf{Improvement Priority} & \textbf{Recommendation Codes} \\
\midrule

\textbf{Interface Clarity \& Interaction Support} &
Comprehensive UI Labeling; Contextual Interface Guidance; Streamlined Data Input; Clear Form Guidance; Enhanced Interface Usability; Interface Modernization; Interface Clutter Reduction; Minimization of Intrusive Ads \\

\addlinespace

\textbf{Navigation, Search \& Information Access} &
Improved Search Discoverability; Enhanced Search Functionality; Simplified Navigational Architecture; Functional Link Maintenance; Enhanced Information Discoverability; Improved Information Clarity \\

\addlinespace

\textbf{System Performance \& Operational Reliability} &
Improved System Performance; Enhanced System Stability; Improved System Scalability \\

\addlinespace

\textbf{Linguistic, Visual \& Cross-Device Accessibility} &
Enhanced Bangla Text Presentation; Native Language Localization; Cross-Device Responsiveness; Improved Visual Contrast \\

\addlinespace

\textbf{Platform Content \& Trust Management} &
Duplicate Listing Moderation; Enhanced Vendor Verification \\

\bottomrule
\end{tabular}
\end{table*}

% \vspace{-0.5em}
\paragraph{Interface Clarity and Interaction Support.}
Participants called for interfaces that were easier to understand and operate. P076 recommended improved labeling and ``tooltips,'' while P101 requested a more ``user friendly input system'' and ``clear instructions'' for completing form fields. Others asked for more user-friendly and modern interfaces; P080 summarized this preference as “make it clean,” while P058 specifically called for fewer intrusive advertisements. Together, these recommendations emphasized reducing interaction complexity while providing clearer guidance.

\paragraph{Navigation, Search, and Information Access.}
Participants also wanted information and functions to be easier to locate. P058 described difficulty finding even the search control and called for a more usable interface, while P075 recommended making ``navigation and finding information'' easier. P067 requested working links and clearer notice boards, and P080 similarly asked for easier navigation. Participants also recommended more effective search, reflecting a broader priority for finding relevant content and actions with less effort.

\paragraph{System Performance and Operational Reliability.}
Performance was a direct priority across several responses. P055 simply recommended that websites ``should be more speedy,'' while P058 called for faster government websites with fewer crashes. P072 similarly recommended supporting ``a large number of clients at a time so that crashes do not occur.'' These suggestions emphasized both immediate performance and the capacity to remain reliable under greater demand.

\paragraph{Linguistic, Visual, and Cross-Device Accessibility.}
Participants identified several improvements aimed at making services accessible across languages, visual conditions, and devices. P076 recommended improving the ``compatibility and aesthetics of Bangla text,'' while P101 called for Bangla translation, particularly for users with limited English proficiency. P067 additionally requested better mobile responsiveness and improved ``color and contrast.'' These recommendations broadened improvement priorities beyond general usability to how services accommodate different users and access conditions.

\paragraph{Platform Content and Trust Management.}
Participants also proposed changes to marketplace content and seller management. P025 recommended preventing multiple sellers from registering the same product and verifying sellers' delivery reliability and whether products maintain the quality described. These suggestions linked easier product selection with greater confidence in marketplace content and sellers.

Overall, these priorities complement the preceding themes by distinguishing what participants \emph{wanted improved} from the barriers they \emph{encountered}. They provide user-identified directions for the broader design and policy implications discussed later.

% \vspace{-0.8em}
\section{Discussion}
\label{sec:discussion}

Our findings reveal that digital accessibility in the Bangladeshi web ecosystem cannot be understood through a single technical or experiential measure. Automated evaluation showed different strengths and limitations across government and non-government platforms, while participants reported a much clearer advantage for non-government services in their everyday interactions. The qualitative findings explain why these perspectives can differ: users' practical access was shaped by barriers that emerged across complete service interactions, from navigating interfaces and completing transactions to obtaining necessary information and relying on the service itself. Together, these findings position technical accessibility as an essential component of broader digital access, but one that does not alone determine whether users can successfully access and use a digital service.

\subsection{Automated Accessibility Does Not Capture the Full User Experience}
\label{sec:discussion_automated}

The variation across our automated measures shows why accessibility should not be inferred from any single technical indicator. Government platforms, for example, had fewer detected WAVE errors and contrast issues and fewer Squidler accessibility and language problems, but this pattern was not consistent across other measures: they produced more WAVE alerts, scored lower on AChecks accessibility and best practices, and showed substantially more Squidler functionality problems. Rather than identifying a technically ``better'' sector, these results reveal different layers of accessibility and web quality that can coexist within the same digital ecosystem. This variation is consistent with prior research showing that automated accessibility tools differ in their evaluation capabilities and the accessibility criteria they cover \cite{fischer2025coverage,ara2025evaluation}. A website may therefore perform relatively well on one class of detectable accessibility problems while still presenting limitations in other aspects of its implementation or functionality.

The clearer sector difference in participants' experiences highlights what these technical measures cannot establish on their own. Participants strongly favored non-government platforms for navigation, usability, clarity, and cross-device experience, while their open-ended accounts described barriers spanning authentication, transactions, error recovery, navigation, and information access. Similar distinctions between technical assessment and experienced accessibility have been examined in other national contexts. Studies in Argentina and Poland, for example, incorporated users' experiences alongside technical assessment to examine barriers that cannot be understood through automated results alone \cite{harari2025argentina,zajac2025poland}. Research on Philippine e-government services has similarly highlighted the role of usability in users' overall service experience \cite{philippines2025usability}. Our findings extend these perspectives to a broader Bangladeshi web-service ecosystem, where comparatively favorable technical indicators did not necessarily correspond to comparatively favorable reports of everyday use.

These experiences often developed across a sequence of interactions rather than through an isolated page element. Automated evaluation is therefore necessary for identifying detectable accessibility problems, but it represents only one layer of practical digital access. Evaluating a digital service also requires examining whether its technical features remain usable and reliable as people move through the broader service process.

% \vspace{-0.em}
\subsection{System Instability and Interaction Friction Limit Practical Access}
\label{sec:discussion_practical_access}

Our findings show that reaching a digital service does not guarantee being able to use it successfully. Theme 1 revealed how loading failures, crashes, authentication problems, transaction failures, and difficulties with error recovery interrupted users at different stages of a task. These difficulties echo problems reported in Bangladeshi hospital websites, where slow or non-loading pages, broken links, and login and appointment-related errors interfered with access to online services \cite{neha2026hospital}. Across our wider range of government and non-government platforms, such disruptions illustrate how the availability of a digital service can be undermined by failures in the interactions required to use it.

Theme 2 revealed a related barrier: unclear navigation, cluttered interfaces, poor information organization, and cross-platform inconsistencies made it more difficult to locate and use available functions. Similar difficulties with slow loading and locating required information have been reported among students using a Bangladeshi public university website \cite{pranty2025usability}. Task-based research on a Chinese government website further showed that search methods, information retrieval, and the length and organization of task paths influenced users' performance \cite{liu2023qingdao}. Together, these findings show that practical access can be limited both by disruptions to task progression and by interaction friction that makes progression unnecessarily difficult.

This distinction is particularly important for transactional and service-oriented platforms, where users often need to move through multiple steps such as authentication, form submission, payment, or booking. WCAG also recognizes the importance of preserving users' progress during interrupted tasks, including allowing them to continue without losing data after re-authentication \cite{w3cwcag}. In our study, participants described how authentication failures, interrupted transactions, and difficulties recovering from errors on Bangladeshi websites disrupted their progress and made it harder to complete intended services. These experiences illustrate how failures in task continuity can turn an otherwise available digital service into a frustrating or incomplete experience. Research on e-government services similarly links interaction-related design characteristics to perceived service quality and continued use \cite{chan2025design}.

Participants' strong preference for simple navigation, effective search, mobile responsiveness, and quick loading further reinforces this point. These qualities address different sources of friction during a task, while system stability supports continued interaction once a task is underway. For service-oriented platforms, evaluation should therefore consider not only whether individual content and functions are accessible, but also whether users can effectively progress through the interactions needed to accomplish their goals.

\subsection{Digital Barriers Can Extend Beyond the Screen}
\label{sec:discussion_service_trust}

Digital access can break down even when users are able to interact with the website itself. Theme 3 showed that incomplete digital processes sometimes required additional offline steps, while missing or unclear information made it harder to obtain the intended service. Similar patterns have been observed in digital government services, where citizens described service access as a multi-channel process involving websites, phone contact, and in-person visits, with additional follow-up creating further burden across the service process \cite{buyannemekh2026burden}. This extends practical access beyond task progression: a digital service must not only support interaction, but also provide the information and service outcomes users seek.

These barriers also intersected with participants' confidence in the services they used. The sector-level differences in perceived security provide quantitative context for the qualitative accounts, where concerns about delays and unofficial costs appeared in experiences with public administrative services, while concerns about reliability and scams appeared in commercial platforms. A study of a digital administrative procedure in Spain similarly found that reliability and security were important for meeting users' expectations, while satisfaction depended on their combination with other aspects of the service experience \cite{catala2025digitalservices}. Experiences with government digital services can also relate to institutional trust: research on government websites in China found that e-government use was associated with trust in government through users' evaluations of government integrity, performance, and responsiveness \cite{li2023egovernmenttrust}. In commercial settings, information confidentiality and integrity have likewise been shown to strengthen consumers' trusting beliefs in e-commerce platforms \cite{singh2024ecommerce}. Together, these findings support our observation that difficulties in accessing information, completing services, or relying on a platform can become connected to perceptions of trust. Accessibility and usability therefore extend beyond the interface to the continuity, information, and reliability of the broader service experience.

\subsection{Accessible Digital Services Require More Than Technical Fixes}
\label{sec:discussion_implications}

Our findings suggest that improving digital accessibility requires attention to both accessible interfaces and the broader conditions that allow users to complete digital services. The priorities identified by participants point toward three complementary areas of action. First, interface clarity, navigation, search, language support, visual accessibility, and cross-device responsiveness should be treated as part of making services easier to find, understand, and use. In the Bangladeshi context, participants' calls for improved Bangla text presentation and native-language localization highlight the importance of providing information in forms that reflect users' linguistic needs. This emphasis is consistent with research showing that accessibility design must account for cultural and linguistic contexts rather than relying solely on generalized design assumptions \cite{nourian2025culture}. Second, performance, stability, and reliable task progression require attention alongside interface-level accessibility, particularly for services involving authentication, forms, payments, or other multi-step interactions. Finally, digital services should support complete online processes, provide clear and accurate service-critical information, and ensure that users can rely on the information needed to complete important tasks. These priorities reflect the need to address both the interactions users encounter and the broader service processes through which their goals are accomplished.

These priorities also have implications for how digital services are evaluated. Automated accessibility testing remains essential, but existing research has shown that automated tools cover only part of accessibility requirements and cannot fully account for users' experiences \cite{fischer2025coverage,ara2025framework}. Our findings therefore support complementing technical evaluation with assessments of complete user tasks and service processes, including how users navigate, recover from failures, access information, and move across devices. Such assessments should examine not only whether users can interact with individual pages, but also whether they can obtain the information and complete the services they need. For government and non-government service providers in Bangladesh, accessibility should therefore be approached as an ongoing responsibility across interface design, system reliability, and service delivery rather than as a static property of individual webpages.

\section{Limitations and Future Work}
\label{sec:limitations}

Some websites could not be successfully evaluated by all three automated tools because of access restrictions, loading failures, or other evaluation difficulties, as also reported in previous large-scale accessibility evaluations \cite{iannuzzi2025combined}. Consequently, tool-specific findings represent the websites successfully evaluated by each tool rather than the complete corpus of 207 websites, and differences across tools could partly reflect differences in the subsets of websites evaluated. The survey primarily represents young, educated, and relatively experienced internet users, limiting the generalizability of the findings to Bangladesh's broader population, particularly older adults, less experienced users, and people with disabilities. Prior research has shown how digital experience can affect older adults' performance in online tasks and how assistive-technology use is shaped by the social and cultural contexts of people with disabilities \cite{taiebmaimon2022older,india2025objectrecognition}. The closed-ended evaluation was also limited to 20 commonly used platforms. Finally, the qualitative findings capture participants' reported experiences rather than directly observed task interactions, limiting our ability to independently verify how individual barriers unfolded during use.

Future work could examine accessibility and usability within specific website categories \cite{webaim2026million}, investigate how technical factors such as contrast, page complexity, and performance relate to user experience, and extend the study through observational evaluations with more diverse users, including people with disabilities. Such evaluations could combine questionnaires and interviews with direct task observation \cite{symptomcheckers2024}, while longitudinal assessments incorporating keyboard and screen-reader walkthroughs could reveal how accessibility changes over time \cite{inal2025longitudinal}.

\section{Conclusion}
\label{sec:conclusion}

This study examined digital accessibility and usability across Bangladesh's government and non-government web ecosystem through automated evaluation of 207 websites and a survey of 103 users. The automated evaluation revealed different strengths and limitations across accessibility, functionality, web practices, and performance measures, while participants reported a clearer preference for non-government platforms in their everyday interactions. Together, these findings show that automated measures provide an important but incomplete view of how digital accessibility is experienced in practice.

Our qualitative analysis further showed how system instability, interaction friction, incomplete digital processes, and information gaps can limit users' ability to progress through digital services and obtain the outcomes they seek. These experiences were also connected to participants' perceptions of service reliability and trust. Digital accessibility should therefore be considered across the broader service process, with attention to accessible and understandable interfaces, reliable task progression, information access, and service continuity. By bringing technical measurements together with users' experiences, this study provides a foundation for designing more accessible and dependable digital services in Bangladesh.

\bibliographystyle{IEEEtran}
\bibliography{references}
\appendix

\section{Squidler Automated Evaluation Platform and Metric Specifications}
\label{sec:appendix_squidler}

This appendix documents the Squidler evaluation process and metrics used during our data-collection period in 2025. The platform has since undergone changes to its interface and evaluation mechanisms. We therefore describe the evaluation process and metric definitions as they applied during the study and analyze the outputs recorded at that time.

During our evaluation period, Squidler performed automated browser-based testing by navigating websites and interacting with page elements. The platform examined functionality, accessibility, and language-related issues across the pages and interactions explored during each scan~\cite{Squidler2024Shopify}. We analyzed the outputs generated by these evaluations as recorded during data collection. The original evaluation reports provided five metrics, which were used in our analysis as follows:

\begin{itemize}
    \item \textbf{Pages:} The number of pages examined during the automated scan, indicating how much of the website was covered during the evaluation.

    \item \textbf{Actions:} The number of interactions performed by the automated agent while navigating and testing the website, reflecting the extent of interaction explored during the scan.

    \item \textbf{Functionality:} The number of functional issues reported during automated interaction, including problems such as broken links, crashes, and dead ends. A lower count indicates fewer functional issues detected within the scope of the scan.

    \item \textbf{Accessibility:} The number of accessibility-related issues reported during the automated evaluation. A lower count indicates fewer detected accessibility issues, but does not imply complete accessibility conformance.

    \item \textbf{Language:} The number of language-related issues reported in the examined content, including spelling and grammar problems. A lower count indicates fewer language issues detected during the scan.
\end{itemize}

The three issue-count metrics were analyzed using the values reported by the platform during our data collection. Pages and Actions were used to describe the scope of the automated evaluation rather than as measures of website quality. These definitions correspond to the Squidler outputs available during the study and were used to interpret the recorded values.

\section{Qualitative Codebook}
\label{app:qualitative_codebook}

Table~\ref{tab:qualitative_codebook} presents the complete codebook used in the qualitative analysis. Refined open codes are grouped under their corresponding axial categories, which are organized by the three themes reported in Section~\ref{sec:qualitative_findings}.

\begin{table*}[t]
\centering
\caption{Codebook showing the mapping of refined open codes to axial categories and overarching themes.}
\label{tab:qualitative_codebook}
\small
\renewcommand{\arraystretch}{1.08}

\begin{tabular}{
    p{0.25\textwidth}
    p{0.32\textwidth}
    p{0.37\textwidth}
}
\toprule
\textbf{Axial Category} &
\textbf{Definition} &
\textbf{Refined Open Codes} \\
\midrule

% ==================== THEME 1 ====================

\multicolumn{3}{l}{
\textbf{Theme 1: System Instability and Barriers to Digital Task Completion}
} \\
\addlinespace[2pt]

\textbf{System Performance, Stability \& Availability} &
Slow, unresponsive, unavailable, or unstable system behavior that delays or interrupts access to a digital service. &
Prolonged System Latency;
Inadequate System Scalability;
Prolonged System Unresponsiveness;
Intermittent Performance Lag;
System Loading Failure;
Time-Dependent Service Unavailability;
System Crash;
Registration Stage System Crash;
Unspecified System Error;
Unhandled Runtime Exception
\\
\addlinespace[4pt]

\textbf{Authentication \& Access Barriers} &
Problems with login, identity verification, authentication, or maintaining access to an active session. &
Redundant Authentication Protocol;
Lack of Session Persistence;
Authentication Latency;
One-Time Password (OTP) System Failure;
Inconsistent Identity Validation
\\
\addlinespace[4pt]

\textbf{Task \& Transaction Workflow Failures} &
Failures within multi-step digital processes that disrupt or prevent tasks such as payments, bookings, or registrations. &
Mobile Payment Interaction Failure;
Erroneous Financial Deduction;
Critical Transaction Failure;
Payment Gateway Loading Failure;
Booking Task Failure;
Onboarding Process Failure
\\
\addlinespace[4pt]

\textbf{System State \& Error Handling Deficits} &
Problems preserving progress or recovering from errors, including conflicting or unresolved system states. &
Delayed Error Recovery;
Restrictive Error Recovery;
System Data Integrity Issue;
State Synchronization Error;
State Loss on Navigation;
Unresolved Account Error State
\\

\midrule

% ==================== THEME 2 ====================

\multicolumn{3}{l}{
\textbf{Theme 2: Interaction Cost and Navigational Friction}
} \\
\addlinespace[2pt]

\textbf{Interface Design \& Interaction Friction} &
Interface-level problems that make controls, layouts, or content harder to understand or interact with efficiently. &
Suboptimal Interface Affordance;
Suboptimal Interface Design;
Excessive Visual Clutter;
Catalog Information Overload;
Excessive Advertisement Exposure;
Detrimental Content Prioritization;
Suboptimal Geolocation Interface
\\
\addlinespace[4pt]

\textbf{Navigational \& Wayfinding Friction} &
Difficulties locating information, understanding where to go next, using search, or navigating poorly organized digital content. &
Ambiguous Navigational Structure;
Erratic Navigational Behavior;
Disjointed Information Architecture;
Poor Information Architecture;
Suboptimal Search Efficacy;
Product Discoverability Friction
\\
\addlinespace[4pt]

\textbf{Cross-Platform \& Device Inconsistency} &
Inconsistent functionality or compatibility across devices or platforms that limits access to the same service. &
Cross-Platform Feature Inconsistency;
Device-Specific Compatibility Failure
\\

\midrule

% ==================== THEME 3 ====================

\multicolumn{3}{l}{
\textbf{Theme 3: Socio-Technical Barriers to Information Access and Service Trust}
} \\
\addlinespace[2pt]

\textbf{Service Fulfillment \& Digital Discontinuity} &
Breaks between online interaction and completion of the intended service, including offline dependencies and fragmented processes. &
Absence of End-to-End Digital Fulfillment;
Offline Escalation Requirement;
Fragmented Administrative Workflow;
Disproportionate Service Friction;
Suboptimal Service Experience
\\
\addlinespace[4pt]

\textbf{Information Clarity \& Visibility Deficits} &
Missing, unclear, or inaccessible information needed to understand, track, or complete a digital service. &
Ambiguous Instructional Content;
Delivery Information Visibility Deficit;
Tracking Functionality Failure;
Online Information Availability Gap;
Content Availability Failure
\\
\addlinespace[4pt]

\textbf{Systemic Trust \& Transparency Deficits} &
Concerns about reliability, transparency, or service practices that reduce users' confidence in depending on a digital service. &
Systemic Trust and Transparency Deficit;
Platform Reliability and Trust Deficit
\\

\bottomrule
\end{tabular}
\end{table*}
\end{document}